\documentclass[a4paper, 12 pt]{article}
\usepackage{amsmath}
\usepackage{amsfonts}
\usepackage{amsthm}
\usepackage{amscd}
\usepackage{amssymb}
\usepackage{graphicx}
\usepackage{color}
\usepackage{graphics}
\newcommand{\myd}{ {\mbox{d}}^{ \uparrow}}
\newcommand{\dpst}{ d_{P}}
\newcommand{\myPhi}{ {{\Phi}}^{ \uparrow}}
\newcommand{\myPhit}{ {{\tilde{\Phi}}}^{ \uparrow}}
\newcommand{\ct}{\tau}
\newcommand{\tp}{t_p}
\newcommand{\tsig}{\sigma^{1t}_p}

\newcommand{\eo}{E_0}
\newcommand{\ec}{E}

\newcommand{\bsigma}{\bar{\sigma}}
\newcommand{\btsig}{\bsigma^{1t}_p}

\newcommand{\dk}{d^1_{\mbox{\large{$\kappa$}}}}
\newcommand{\dkk}{d^2_{\mbox{\large{$\kappa$}}}}
\newcommand{\hk}{H^{\mbox{\large{$\kappa$}}_1}_{\Phi_0,\Phi_1}}
\newcommand{\hkk}{H^{\mbox{\large{$\kappa$}}_2}_{\Phi_0,\Phi_1}}
\newcommand{\hkt}{\tilde{H}^{\mbox{\large{$\kappa$}}_1}_{\Phi_0,\Phi_1}}
\newcommand{\hkup}{H^{\uparrow\mbox{\large{$\kappa$}}_1}_{\Phi_0,\Phi_1}}
\newcommand{\hkupt}{\tilde{H}^{\uparrow\mbox{\large{$\kappa$}}_1}_{\Phi_0,\Phi_1}}

\newcommand{\half}{{\textstyle \frac{1}{2}}}

\newcommand{\qqh}{\hat{C}}
\newcommand{\qq}{{{C}}}

\newcommand{\Phat}{{\hat P}}

\newcommand{\Rr}{{\mathbb R}}

\newcommand{\dd}{\mbox{\huge .}}

\newtheorem{thm}{Theorem}[section]
\newtheorem{lem}{Lemma}[section]
\newtheorem{prop}{Proposition}[section]

\theoremstyle{definition}
\newtheorem{defn}{Definition}[section]

\theoremstyle{remark}
\newtheorem{rem}{Remark}[section]

\font\fontr=msbm10 scaled 1400
\newcommand{ \dst} {\displaystyle}

\newcommand{\eps} {\epsilon}

\newcommand{\lb} {\left(}
\newcommand{\rb} {\right)}
\newcommand{\lbr} {\left\{}
\newcommand{\rbr} {\right\}}

\newcommand{\al} {\alpha}

\newcommand{\be} {\beta}

\newcommand{\ga} {\gamma}
\newcommand{\bga}{\begin{array}{l}}
\newcommand{\ena}{\end{array}}
\newcommand{\bge}{\begin{equation}}
\newcommand{\ene}{\end{equation}}
\newcommand{\kk}{\mbox{\Large{$\kappa$}}}
\newcommand{\R} {\mbox{\fontr{R}}}

\def\comment#1{}

\def\withcomments{
\addtolength{\oddsidemargin}{-0.5 in}
\addtolength{\evensidemargin}{-0.5 in}
\newcounter{mycommentcounter}
\def\comment##1{\refstepcounter{mycommentcounter}%
  \ifhmode%
  \unskip%
  {\dimen1=\baselineskip \divide\dimen1 by 2 %
    \raise\dimen1\llap{\tiny -\themycommentcounter-}}\fi%
  \marginpar{\renewcommand{\baselinestretch}{0.8}%
    \footnotesize [\themycommentcounter]: \raggedright ##1}}
}

\withcomments

\begin{document}

\date{}
\title{Homotopy classification of director fields on polyhedral domains with tangent and periodic boundary conditions
, with applications to bi-stable post-aligned liquid crystal displays.
}
\author{M. Zyskin
%\thanks{M.Zyskin@bristol.ac.uk}
\\
%Department of Mathematics, University of Bristol,  BS8 1TW, UK.
%456 Washington Ave, Apt 3K, Belleville, NJ 07109.
}

\thispagestyle{empty}

\maketitle

\begin{abstract}
%We investigate   a technique to study homotopy classification problems for maps of cell complexes to a topological space, with subcomplexes required to be mapped to subspaces. We  obtain homotopy classification for  problems which have applications to  nematic liquid crystal in polyhedral geometry: classification of director fields inside a  simply-connected aspherical polyhedron in $\R^3 $ with tangent boundary conditions on faces; and  in a  region outside such polyhedron, with tangent boundary conditions on faces of polyhedron and   fixed-degree or  periodic external boundary conditions.

We obtain complete topological classification of inequivalent states of nematic liquid crystal in the geometry of periodic array of rectangular posts between two parallel slabs, with tangent or normal boundary conditions. This classification has applications in bi-stable pos-aligned liquid crystal display design and have technological significance. Methods used in classification are those of algebraic topology and go beyond relative homotopy groups. 

\end{abstract}

\section{Introduction}
This paper introduces  an   approach in the style of Eilenberg \cite{eilenberg} to study homotopy classification problems which appear  in the context of topological classification of states of nematic liquid crystal in polyhedral domain with tangent or a mixture of tangent and periodic boundary conditions. 

Topological classification of liquid crystal
configurations in $\Rr^3$ as well as in domains with smooth boundary and Dirichlet boundary conditions
has been extensively studied -- see, eg, Mermin~\cite{mermin}, de
Gennes and Prost~\cite{degennes}, and Kl\' eman~\cite{kleman}. For domains with smooth boundary with  tangent or normal boundary conditions the local classification of  defects is equivalent to computing relative homotopy groups  \cite{loc}. However this is not sufficient for the  classification of defects at vertices (or edges) of a polyhedron. We are also interested in the global  classification of maps up to homotopy 
%of possible nematic states  
rather then in just local classification of defects. 
For the case of nematic liquid crystal in  polyhedral cells $P\subset\R^3$ with tangent boundary conditions on
faces, topologically inequivalent continuous configurations of nematic liquid crystal are described by homotopy classes of director fields (i.e. maps $P\rightarrow RP^2$), which on faces of $P$ must be parallel to the face. Related  homotopy classification of tangent maps  $P\rightarrow S^2$ in the case when $P$ is   convex was obtained  in \cite{rz} using explicit construction of homotopies (which is very tedious).   Homotopy classification of \cite{rz} was subsequently used to analyze  harmonic map problems in fixed homotopy classes \cite{mz}- \cite{mrz4} (for the case of  point defects or holes in $\Rr^3$ with fixed degree, such harmonic map problems were studied in \cite{bli}).

The paper is organized as follows. In section 2 we introduce  the necessary background and notations. We define  similarly to \cite{eilenberg}  certain homotopy invariants of maps and pairs of maps  on cells   with values  in appropriate homotopy groups (Definitions 2.1 and 2.3), and describe their  basic properties and basic  extension and homotopy classification results.
%In section 3, we show using those basic extension and classification results that 2  maps from a simply-connected  polyhedron $P$ (not containing any non-contractable two-spheres) to $RP^2$, which on faces of  $P$ are required to be tangent to the face, are homotopic if and only if invariants of pair of maps associated to cleaved edges of polyhedron (edges introduced by removing little 3-cells near vertices) are zero, and   invariants of a certain map associated to cleaved faces (obtained by gluing the pair of maps using homotopies on edges) are zero (Theorem 3.1). It also establishes sum rules for the above invariants (Proposition 3.3).

In section 3 we obtain homotopy classification of director fields on a contractible polyhedron $P$, with tangent boundary conditions on faces, using homotopy invariants and basic extension properties, but not explicit homotopies.  Tangent boundary conditions on faces imply that on  edges  of $P$ director field must be parallel to the edge, and thus it must be discontinuous at vertices of $P$; we assume that it is continuous elsewhere.  Homotopy classification of such discontinuous director fields on $P$ is equivalent to classification of {\em continuous} director fields on a  {\em truncated polyhedron} $\Phat$ (Section 2.1). 
%We define truncated polyhedron $\Phat$ by removing from $P$ small  3-balls centered at vertices of $P$. Faces and edges of truncated polyhedron are of two kinds, truncated faces and edges (parts of faces and edges of $P$), and cleaved faces and edges introduced by excised small balls near vertices. Homotopy classification of tangent director fields on $P$ is equivalent to homotopy classification of director fields on $\Phat$ with tangent boundary conditions on truncated faces (there is no restriction on cleaved faces).
%, apart from continuity).

For a pair of tangent maps $\Phi_0, \Phi_1$, we define homotopy invariants  associated to   edges of $\Phat$ with values in $\pi_1 (RP^1)$, and use those invariants as coefficients in the formal chain $d^1 (\Phi_0, \Phi_1) $ (Eq. (\ref{d1})). 
%The restrictions of  $\Phi_0$ and $\Phi_1$ to edges of $\Phat$ are homotopic  if and only if  $d^1 (\Phi_0, \Phi_1)=0$.
If  $d^1 (\Phi_0, \Phi_1)=0$, there exist  homotopies on   edges of $\Phat$ satisfying tangent boundary conditions, moreover we can construct certain homotopy invariants 
with values in $\pi_2 (S^2)$, 
associated to  faces of $\Phat$, using  homotopies on edges to glue along the edges the restrictions of maps $\Phi_0, \Phi_1$ to faces.
%(and lifting all the maps to $S^2$, the universal covering space of $RP^2$).
%, moreover, using  
%Using those tangent homotopies and maps $\Phi_0, \Phi_1$  and lifting all the maps to maps to the universal covering space,  
We show that such invariants do not depend from a choice of tangent edge homotopies, and only depend from $\Phi_0, \Phi_1$ (Proposition \ref{Prop:SurfaceHomConvex}).
%Using those tangent homotopies and maps $\Phi_0, \Phi_1$  and lifting all the maps to maps to the universal covering space,  we construct certain homotopy invariants with values in $\pi_2 (S^2)$, associated to  faces of $\Phat$, and show that those invariants do not depend from a choice of edge homotopies, and only depend from $\Phi_0, \Phi_1$ (Proposition \ref{Prop:SurfaceHomConvex}). 
Thus  if  $d^1 (\Phi_0, \Phi_1)=0$, we can define a formal chain
$d^2 (\Phi_0, \Phi_1)$ (\ref{d2}) with the above invariants as coefficients. 
We show (Theorem \ref{ThmConvexHom}) that $\Phi_0, \Phi_1$ are homotopic if and only if $d^1 (\Phi_0, \Phi_1)=0$ and
$d^2 (\Phi_0, \Phi_1)= 0$. 

Results of section 3 were first obtained for convex polyhedra in \cite{rz}, using an equivalent formulation in terms of unit-vector fields and  explicit homotopies construction.  Basic extension and classification results allow to streamline homotopy classification, generalize  it immediately  to any contractible polyhedron, not just convex one, and  apply the technique to study more complicated classification problems.
%, such as the one analyzed in Section 4 (a problem with significant technological applications)
Advantage of using explicit homotopies  \cite{rz} is that 
%as a by-product 
it provides representative maps in each homotopy class.
%re-derive results of \cite{rz} without using explicit homotopies, and only using homotopy invariants and extension properties.

We illustrate our method by  providing homotopy classification of   nematic liquid crystal states in the domain between two horizontal plates  outside a periodic array of rectangular posts located on the bottom plate (section 4). Boundary conditions are tangent on the bottom plate and the post surface, and  periodic with respect to integer translations in horizontal directions, with   normal or  tangent  boundary conditions on the top plate. This problem is important due to applications in new  bi-stable liquid crystal displays \cite{hp}. In \cite{mrz5} configurations of nematic of several simple topological types  were studied in such geometry, and energy-minimising configurations were found numerically to be smooth away from the vertices.  In this paper  we provide complete topological classification of states continuous away from the vertices in such geometry (Theorem (\ref{ThmPostHom}). %Such classification enables us to investigate all topologically distinct local energy minima, not just those of simple topology. 

In the paper we assume that director fields are as continuous as possible, so they are continuous except  at vertices where  tangent boundary conditions disallow  continuity. More general types of singularities can be classified using similar technique. %We note that  \cite{rz} was using the fact that on convex polyhedron director field can be lifted to a unit-vector field. That is no longer true when polyhedron is not simply-connected, or in the case of periodic boundary conditions. In this paper we work with director fields.
%The new approach allows to deal with unit-vector or director fields in unified fashion, as  the target space  appears in all the constructions only via its homotopy groups.

In this paper we are focusing on homotopy classification problems for nematic liquid crystals. However,  similar technique may  be used algorithmically for  boundary-value problems for other ordered media, such as spin systems, superfluid helium-3, etc. Indeed, in our approach  all the relevant information about the target space is contained in its homotopy groups. 

{\bf Notations.} Throughout the paper, we use  symbol $\simeq$ to denote  homotopic maps. We use symbol $\cong$ to denote homeomorphic spaces. If $\Phi: X \rightarrow Y$  and $W$ is the universal covering space for $Y$, we denote by  $\myPhi$ the lifted map $X\rightarrow W$; such lifted map is determined by lifting at one point. In our examples motivated by nematic liquid crystals, $Y=RP^2$, and the universal covering space is $S^2$. We denote by $\eps_{\al \be}$  the antisymmetric tensor with $\eps_{12}=1$.
\section{
%Cell complexes,
%Generalized degrees  and
Homotopy Invariants, Extension and Classification
%and $\lambda_c$ homotopy.
}
%cell complex
%In the paper we will be concerned with maps of polyhedron, with tangent
In the paper we will be concerned with homotopy classification of
%reff Req
 maps of a polyhedron to $RP^2$, describing topologically inequivalent states of nematic liquid crystal, with tangent or periodic boundary conditions on faces. In
%the beginning
this section
we introduce some notations, and briefly outline a more general set-up in which it is natural  to address such classification problems. We largely follow notations and terminology of Eilenberg paper \cite{eilenberg}, which is simple, very readable, and is straight to the point. A physicist-oriented introduction to the subject, with lots of pictures drawn,  is an excellent review by Mermin \cite{mermin}.  More pedantic definitions of the main objects can be found e.g. in \cite{rokh}. %Unfortunately those tools are not standard among applied mathematicians (although in our view they should be).

%!! Finite cell complex, see e.g. 
% We briefly describe it here. We use open cells, as in    They are homeomorphic to

An open   k-cell $\sigma^k$ in   $\R^N$
is a subspace homeomorphic to an open k-disk.
A finite cell complex $X$ is a space  (we 
may
consider it to be a subspace of $\R^N$ for large enough $N$), which can be partitioned into finitely many cells, so that for every cell $\sigma^k \subset X$ there is a continuous map $\chi_{\sigma^k}$
%(called characteristic map)
of a closed k-disk $D^k$ into  $X$, which maps interior of $D^k$ homeomorphically onto $\sigma^k$, and maps $S^{k-1}$, the boundary of  $D^k$  onto a union of cells of dimension less then $k$. If $K$ is a cell complex,  the n-skeleton  $K_n$ is  a subcomplex consisting of all cells of dimension $\leq n.$

We denote by $\ec$ the closed interval $[0,1]$. For a set $A$, we denote by $\bar{A}$ the closure of $A$, and by $A_0$ the interior of $A$
%(e.g.  $\bsigma^k$ is the closure of a cell $\sigma^k$, $\eo$ is the open interval $(0,1)$, etc.)
.

We denote by $\lb{\sigma^k}\rb^{\dd}$ the $(k-1)-$ sphere parameterizing the boundary of a cell $\sigma^k .$ 
%(if the boundary of $\sigma^k$ is homeomorphic to $S^{k-1}$ we do not distinguish between $\lb{\sigma^k}\rb^{\dd}$  and the boundary).  
Given a map $\Phi: X\rightarrow Y$, its restriction to $\lb{\sigma^k}\rb^{\dd}$, denoted $\Phi\vert \lb{\sigma^k}\rb^{\dd}$ is a map $ \Phi \circ \chi_{\sigma^k}: S^{k-1} \rightarrow Y$. 
%nq
If $W\subset X$ is such that its interior $W_0 $ is a cell $\sigma^k$, by  $\lb W\rb^{\dd}$ we mean $\lb W_0 \rb^{\dd}$.

Two maps $\Phi_0, \Phi_1: S^n \rightarrow Y$ are (free) homotopic, if there is a continuous map $\Phi: S^n \times E \rightarrow Y$, 
such that $\Phi\vert \lb S^n\times \{ m \} \rb = \Phi_m$, $m=0,1$. Given basepoints $s_0\in S^n, y_0\in Y$, two  maps $\Phi_0, \Phi_1: S^n \rightarrow Y$,  $\Phi_0(s_0) = \Phi_1(s_0)=y_0$  are   homotopic maps between spaces with basepoints $(S^n,s_0)\rightarrow (Y, y_0)$ if there is a map $\Phi: S^n \times E \rightarrow Y$, such that $\Phi\vert \lb S^n\times \{ m \} \rb   = \Phi_m, \ m=0,1$  and   $\Phi\vert \lb \lbr s_0\rbr \times E \rb =y_0$.  Homotopy classes of maps  $(S^n,s_0)\rightarrow (Y, y_0)$ is  a group $\pi_n (Y,y_0)$.
A topological space $Y$ is called $n$-simple, if   any free homotopy class of maps $S^n\rightarrow Y$ contains  exactly one homotopy class of maps between spaces with  basepoints $(S^n,s_0)\rightarrow (Y, y_0)$. 
%$Y$ is $1$-simple if $\pi_1$ is abelian; in general $Y$ is $n$-simple if the action of $\pi_1 (Y,y_0)$ on $\pi_n (Y,y_0) $ is trivial. (If maps $(S^n,s_0)\rightarrow (Y, y_0)$ are represented by maps of $n$- disk $D^n$ of radius 1 to $Y$, with boundary of $D^n$ mapped to $y_0$, the action of $\pi_1 (Y,y_0)$ on    $\pi_n (Y,y_0) $ is obtained by taking a radially rescaled  representative map in  $\pi_n (Y,y_0) $ on a disk of radius $1/2$, followed by a representative path in $\pi_1 (Y,y_0)$ in every radial direction on a spherical ring with inner radius $1/2$ and outer radius $1$). 
If $Y$ is $n$-simple we can write $\pi_n (Y)$ unambiguously to describe free homotopy classes $S^n\rightarrow Y$ regardless of a basepoint.
 
\begin{figure}[h]
\begin{center}
\input{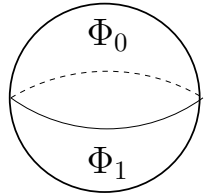}
\caption{$(\Phi_0, \Phi_1,\sigma^k, Y)$ }
\label{fig1:()}
\end{center}
\end{figure}

\begin{defn}
Let  $\sigma^k$ be a k-cell,
%$Y_k$ a subspace of
$Y$
a $k$-simple
topological space,  and    $\Phi_0, \Phi_1   $ maps $\bsigma^k \rightarrow Y$ which agree on the boundary,
%$(\sigma^k)^{\dd}$
$\Phi_0\vert (\sigma^k)^{\dd} = \Phi_1\vert (\sigma^k)^{\dd} .$ Let $h_+$  be  a homeomorphisms of the closed  
%reff closed
northern  hemisphere  of $S^k$ to a disk $D^k,$   preserving  the orientation, and $h_-$  a homeomorphism  of the closed southern  hemisphere  of $S^k$ to $D^k,$    reversing  the orientation, and agreeing with $h_+$ on the equator. 
Let
%\bge
$
(\Phi_0, \Phi_1,\sigma^k, Y)
$
%\label{def:()}
%\ene
be the map  $S^k \rightarrow Y$  given by $\Phi_0 \circ \chi_{\sigma^k} \circ h_+$ in the northern hemisphere, and by $\Phi_1 \circ \chi_{\sigma^k}\circ h_-$ in the southern hemisphere, see Figure \ref{fig1:()}.
%\noindent
%If  $Y$ is  $k$-simple, 
We   define
\bge
%$
d (\Phi_0, \Phi_1,\sigma^k, Y)
%$
\label{d(sigma)}
\ene
to be the element of $\pi_k (Y),$ corresponding to the map $(\Phi_0, \Phi_1,\sigma^k, Y).$ 
%%!!!do we need it?
If $W_0 $, the interior of $W\subset X$, is a cell $\sigma^k$,  by
%$(\Phi_0, \Phi_1, W, Y) $ as  $(\Phi_0, \Phi_1, W_0, Y)$, and 
$d (\Phi_0, \Phi_1,W, Y)$ we mean $ d(\Phi_0, \Phi_1, W_0, Y)$.
\end{defn}
%\noindent Note. 

\label{def:d}
%Let $\Phi_0, \Phi_1: K\rightarrow Y,$ and $\Phi_0\vert K_{k-1} = \Phi_1\vert K_{k-1}.$
%\end{defn}

\begin{defn}
Let  $\sigma^k$ be a k-cell,
$Y$ a $(k-1)$-simple topological space,  and    $\Phi$ a map   \mbox{$(\sigma^k)^{\dd} \rightarrow Y$.}
%defined on the boundary of $\sigma^k$. 
%$(\sigma^k)^{\dd}$ is homeomorphic to $S^{k-1}.$
We denote by $c^k(\Phi, \sigma^k, Y)$ the  element of $\pi_{k-1} (Y)$ corresponding to $\Phi\vert(\sigma^k)^{\dd}.$ 
%If $W$ is a set such that  its interior $W_0$ is a cell  $\sigma^k$,
If $W_0 $, the interior of $W\subset X$, is a cell $\sigma^k$, we define $c^k(\Phi, W, Y)$ as  $c^k(\Phi, W_0, Y)$ . 
%In this paper, we will be often concerned with $Y=RP^2, $ and
\label{def:c}
\end{defn}
%\noindent Note. 
%\vspace{4mm}

%We  will use notation $c^k(\Phi, \sigma^k)$ for $c^k(\Phi, \sigma^k, RP^2),$ $k=2 , 4$.  Given a map $\Phi: (\sigma^3)^{\dd} \rightarrow RP^2$, we denote by  $c^3(\Phi, \sigma^3)$   elements of $\pi_{2} (S^2)$ corresponding to the lifted to the universal covering space map   $\myPhi : ((\sigma^3)^{\dd},x_0)  \rightarrow (S^2, s_0) $ (by a choice of lifting at a basepoint, this is defined up to a sign  when this element $\pi_{2} (S^2)$ is nonzero, and uniquely defined if it is zero. In what follows, we will encounter conditions $c^3(\Phi, \sigma^3) = 0$; this is unambiguously defined regardless of the choice of lifting)

%\begin{rem}
Let us cut  $(k-1)$-sphere $(\sigma^k)^{\dd}$ into two disks $D^{k-1}_+$, $D^{k-1}_-$. Let $h$ be a homeomorphism from $D^{k-1}_+$ to $D^{k-1}_-$, which is identity on the boundary of $ D^{k-1}_+$. 
%Let $\sigma^{k-1}_+$ be the interior of $ D^{k-1}_+$. 
It's clear that
\bge
c^k(\Phi, \sigma^k, Y)= d(\Phi\vert D^{k-1}_+, \Phi\vert D^{k-1}_- \circ h  , D^{k-1}_+, Y).
\label{cISd}
\ene
%\end{rem}

In the paper  we are concerned with $Y=RP^2$, since 
mean direction of molecules of a nematic liquid crystal
%at a point in physical space is described by a director, a unit vector with opposite directions identified. A director is
is described by a director, that is a point in $RP^2$.  $RP^2$  is $n$-simple for $n=1, 3$, but not for $n=2$ (nontrivial element of $\pi_1(RP^2)$ acts as $(-1)$ on  $\pi_2(RP^2)$). Thus $d (\Phi_0, \Phi_1,\sigma^k, RP^2)$ is defined by (\ref{d(sigma)}) for $k=1,3$ but not for $k=2$, and  $c^k(\Phi, \sigma^k, RP^2)$  is defined by Definition \ref{def:c}
for $k=2,4$, but not for $k=3$. We  resolve this difficulty by lifting a map $S^2 \rightarrow RP^2$ to  a map  to  $S^2$,  the universal covering space for $RP^2$, which  is $n$-simple for any $n$.  
Indeed, by the covering homotopy lemma (see eg \cite{novikov}), a map $\phi: X\rightarrow RP^2$ with $\pi_0 (X)=\pi_1 (X) = 0$ (in particular $X=S^2$) can be lifted to a map ${\phi}^{\uparrow}: X\rightarrow S^2$.   Such lifting is  uniquely determined by lifting at one point $x_0\in X$. For paths in $X$ starting at $x_0$ the lifting is then defined by continuity; since all paths in $X$ with fixed endpoints are homotopic, this gives a consistent lifting at all other points of $X$. Since the projection from  $S^2$ to $RP^2$ (
%which can be represented by 
identifying diametrically opposite points of $S^2$) is a 2:1 map, there are two ways of lifting the map at one point. Given a map $(\Phi_0, \Phi_1,\sigma^2, RP^2): S^2 \rightarrow RP^2$,  we lift it to a map
$S^2 \rightarrow S^2$ by choosing one of the two possible liftings at one point,  
%$x_0\in S^2 \mapsto s_0 \in S^2$, 
and denote by
%$d(\Phi_0, \Phi_1; \sigma^2, x_0; S^2, s_0)$
\bge
 \myd (\Phi_0, \Phi_1, \sigma^2, S^2)\equiv d (\myPhi_0, \myPhi_1, \sigma^2, S^2)
 \label{dLift}
\ene
the element of $\pi_2 (S^2)$ corresponding to the lifted map.   

Similarly, given a map $\Phi\vert(\sigma^3)^{\dd}: S^2  \rightarrow RP^2$, we lift it to a map $\myPhi\vert(\sigma^3)^{\dd}: S^2  \rightarrow S^2$. %and   denote by  
$c^3(\myPhi, \sigma^3, S^2)$ is the  element  of $\pi_{2} (S^2)$ corresponding to the lifted  map. 

Due to  the choice of   lifting at one point, $\myd (\Phi_0, \Phi_1, \sigma^2, S^2)$ , $c^3(\myPhi, \sigma^3, S^2)$ are in general defined up to a sign, but are uniquely defined if those elements are zero. 
%\end{rem} 
%Note that due to choice of lifting at one point, $c^3(\myPhi, \sigma^3, S^2)$   is defined up to a sign  when that element of $\pi_{2} (S^2)$ is nonzero, but is uniquely defined if it is zero.

%\vspace{3mm}

%\begin{defn}
Let $A$ be a subspace of $X.$ We recall that maps $\Phi_0$, $\Phi_1:$ $X\rightarrow Y,$ $\Phi_0\vert A = \Phi_1\vert A$ are homotopic relative to $A $ (denoted $\Phi_0 \simeq \Phi_1 \ rel \ A$) if there is $\Phi: X\times E \rightarrow Y,$ 
such that
$$
\bga
\Phi\vert X\times \{i\}= \Phi_i , \quad i=0,1;
\\
\Phi\vert A\times E = \Phi_0\vert A =\Phi_1\vert A \  .
\ena
$$
%\end{defn}
\begin{prop}
{\mbox{}
\\
1. a) Let $\Phi_0, \Phi_1: \bar{\sigma}^k\rightarrow Y$, where $Y$ is $k$-simple (e.g. $Y=RP^2$, $k=1,3$),  and $\Phi_0\vert(\sigma^k)^{\dd} = \Phi_1\vert(\sigma^k)^{\dd}$ .
Then
\bge
\Phi_0 \simeq \Phi_1 \ rel \ (\sigma^k)^{\dd} \Leftrightarrow d (\Phi_0, \Phi_1,\sigma^k, Y) =0.
\label{relHomViaD}
\ene
\\
b) Let $\Phi_0, \Phi_1: \bar{\sigma}^2\rightarrow RP^2$ and  $\Phi_0\vert(\sigma^2)^{\dd} = \Phi_1\vert(\sigma^2)^{\dd}$ .
Then $$\Phi_0 \simeq \Phi_1 \ rel \ (\sigma^2)^{\dd} \Leftrightarrow d (\myPhi_0, \myPhi_1,\sigma^2, S^2) =0,$$ where $\myPhi_0, \myPhi_1$ are lifted maps to $S^2$, such that  $\myPhi_0\vert(\sigma^2)^{\dd} = \myPhi_1\vert(\sigma^2)^{\dd}$
%(the latter condition will be satisfied    if  we use the same lifting  $\myPhi_0, \myPhi_1$   at a point  on  $(\sigma^2)^{\dd}$)
.
\\
2. Let $\Phi_0, \Phi_1, \Phi_2  $ be   maps $\bar{\sigma}^k \rightarrow Y,$   $\Phi_0\vert (\sigma^k)^{\dd} = \Phi_1\vert (\sigma^k)^{\dd} = \Phi_2\vert (\sigma^k)^{\dd}$, and let $Y$ be $k$-simple %(we will need this for $k=1, 3$ for $Y=RP^2$, or $k=2$ and $Y=S^2$)
Then }
\bge
d (\Phi_0, \Phi_1,\sigma^k, Y)= - d (\Phi_1, \Phi_0,\sigma^k, Y),
\label{d:scew}
\ene
\bge
d (\Phi_0, \Phi_1,\sigma^k, Y) +  d (\Phi_1, \Phi_2,\sigma^k, Y)= d (\Phi_0, \Phi_2,\sigma^k, Y).
\label{d:addition}
\ene
\label{prop:RelHomCell}
\end{prop}
The proof is standard and can be found e.g. in \cite{eilenberg}.

\section{Homotopy Classification of Tangent Director Fields on   Polyhedra}
%In \cite{rz}  it is shown (by constructing  explicit homotopies), that  continuous tangent unit-vector fields on convex polyhedron  are classified by homotopy invariants (edge signs, kink numbers, and wrapping numbers, subject to sum rules).  In  this  section  we re-derive results of \cite{rz} using algebraic topology framework \cite{eilenberg}.

%This sets up notations and technique which   simplifies computations in more complicated problems, e.g. involving periodic as well as tangent boundary conditions, considered in ...
%{\bf Tangent director fields on polyhedra.}
\subsection{Tangent director fields on polyhedra.}
Let  $\bar{P}\subset \R^3$ be  a 
%simply-connected  polyhedron not containing any non-contractible 2-spheres (e.g. a convex polyhedron).  
contractible polyhedron, $K_{sing}$ the set of vertices of $\bar{P},$  and  $P= \bar{P}\setminus K_{sing}.$
%Let $Y =RP^2,$ which we model by a unit sphere $S^2\subset \R^3$ with diametrically opposite points identified.
We consider  
continuous 
maps $\Phi: P  \rightarrow Y=RP^2$,   director fields  on $P$.  %We call  points in $RP^2$ {\em directors,} %and identify them with lines in $\R^3$ through the origin or alternatively with points on a unit sphere $S^2\subset \R^3$  with diametrically opposite points identified,
%and maps $\Phi$   
We associate to faces $\sigma^2_i$ of $\bar{P}$ subspaces $Y_i^1\cong RP^1=S^1  \subset RP^2$ of directors  parallel to the face $\sigma^2_i$.
%homeomorphic to $RP^1=S^1$, corresponding to  .
%($Y_i^1$  is the projection of a great circle of $S^2$ parallel to the $i$-th face to $RP^2$, identifying diametrically opposite points of $S^2$). 
We call $\Phi$ a tangent map if  the   restriction of $\Phi$ to any     face  $\sigma^2_i$ of $P$ is  a map to the subspace $Y_i^1.$
%$\Phi(\sigma^2_i)\subset Y_i,$ where $Y_i $ is a subspace of $RP^2$ which to those lines through the origin in  those lines through the origin We call such maps tangent.  map  is to the great circle $S^1_i$ of $S^2$, parallel to the face $\sigma^1_i,$ (so that  the unit-vector field corresponding to such map is tangent to the face.) Let $Y^1_i$ be a projection of that circle to $RP^2, $obtained by identifying diametrically opposite points on $S^2$  ($Y^1_i$ is homeomorphic to $S^1.$ ) It follows from continuity that an edges of $P,$ incident with faces $\sigma^2_i$ and $\sigma^2_j$ must be mapped to the  point $Y^1_i \bigcap Y^1_j$ of $Y.$

%We note that 
%continuous 
Tangent maps  $P\rightarrow RP^2 $   can be lifted to maps to the universal covering space $S^2,$ such that restrictions of the maps to any face of $P$ is a map to the great circle  of $S^2$ parallel to the face.

%\newpage

\begin{figure}[h]
\begin{center}
\input{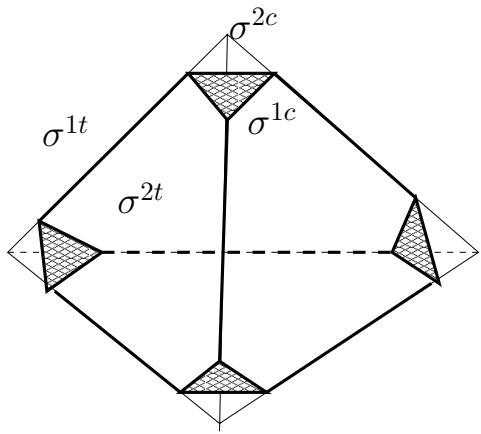}
\caption{Truncated polyhedron $\Phat$.  Cleaved edges $\sigma^{1c}$, faces $\sigma^{2c}$ correspond to   dark regions, while truncated edges $\sigma^{1t}$, faces $\sigma^{2t}$ correspond to white regions.}
\label{fig3:Phat}
\end{center}
\end{figure}

\noindent{\bf Truncated polyhedron $\Phat $
%\cite{rz}.
}
Truncated polyhedron $\Phat $ is obtained by removing from $\bar{P}$  small 3-cells incident to vertices of $\bar{P}$.  For a vertex   $v_i$   of $\bar{P},$    let $b_i =B_{v_i, \epsilon_i} \cap \bar{P}$
%be a 3-cell so that $$
where $B_{v_i, \epsilon_i}$ is a ball with center at  $v_i,$ and with radius $\epsilon_i$ small enough so that
$b_i$ does not intersect any edges or faces not incident to the vertex $v_i$, and $b_i \cap b_j =  \varnothing, i\neq j$. Then $\Phat = \overline{\bar{P}\setminus \cup_i b_i}.$
We view $\Phat$  as a cell complex.
0-cells $\Phat_0$ in this complex are vertices of $\Phat.$
1-cells
%$\Phat_1$
are   edges of $\Phat;$ they are of two kind,
truncated edges $\sigma^{1t}$, which are parts of edges of $P,$
%$ \lbr  \sigma^{1\tau}_i\rbr  \subset \lbr \mbox{  edges of} \  P\rbr $ ,
and cleaved edges $\sigma^{1c}$, introduced by removing 3-cells around vertices.
%$\{ \sigma^{1c}_i\}\subset \Phat \bigcap  \lbr  \mbox{  edges of} \      \bigcup_j b^{3}_j  \rbr.$
2-cells
%$\Phat_2$
are   faces of $\Phat;$ they are of two kind:
truncated faces $\sigma^{2t}$, which are parts of faces of $P,$
%$ \lbr  \sigma^{12t}_i\rbr  \subset \lbr \mbox{  faces of} \  P\rbr $ ,
and cleaved faces $\sigma^{2c}$, introduced by removing 3-cells around vertices (Figure \ref{fig3:Phat}).
% $\{ \sigma^{1c}_i\}\subset \Phat \bigcap  \lbr  \mbox{  faces of} \      \bigcup_j c^{3}_j  \rbr.$
%$\{ \sigma^{2\tau}\}$ (faces of $P$), and cleaved faces $\{ \sigma^{2c}\}.$
There is just one 3-cell,   $\Phat$  itself. A map $\Phat \rightarrow RP^2$ is called tangent, if it is tangent on truncated  faces of $\Phat $ (there is no restriction on cleaved faces).

\begin{prop}
%\prp
{ Two tangent maps $\Phi_0,$ $\Phi_1: P \rightarrow RP^2$  are homotopic if and only if
%tangent maps
%$\Phi_0\vert\Phat,$ $\Phi_1\vert\Phat$
their restrictions to $\Phat$ are homotopic.}
\label{prop:TruncatedOK}
\end{prop}
\noindent For unit-vector fields, it is shown in Proposition 2.1 in \cite{rz}. Generalization to director fields is straightforward.

\subsection{Homotopy Classification}
Let $\Phi_0,$ $\Phi_1$ be two tangent maps  $\Phat \rightarrow Y=RP^2.$ Such  maps necessarily agree on   truncated edges, and on vertices of  $\Phat$, $\Phi_0\vert \Phat_0 = \Phi_1\vert \Phat_0,$ due to tangent boundary conditions.
%since due to  tangent boundary conditions a truncated edge is mapped to the director parallel to the edge. It follows that they  must also agree on vertices of $\Phat,$ $\Phi_0\vert \Phat_0 = \Phi_1\vert \Phat_0$.  
%(as   vertices  of $\Phat$ are  incident to  {  truncated} edges).
Let $\sigma^{1c}_i$ be a cleaved edge, and let $\sigma^{2\tau}_{s(i)}$ denote the truncated face incident to $\sigma^{1c}_i$.
Since $\Phi_0\vert \Phat_0 = \Phi_1\vert \Phat_0,$
%$\Phi_0 $ and $\Phi_1$ agree on vertices of $\Phat$
%)
we can define 
%a map  $ (  \Phi_0, \Phi_1,\sigma^{1c}_i,Y^1_{s(i)} ):  S^1 \rightarrow Y^1_{s(i)}.$  Let
$d(  \Phi_0, \Phi_1,\sigma^{1c}_i,Y^1_{s(i)}) \in \pi_1 (Y^1_{s(i)} )$
%be the corresponding element of
%$\pi_1 (Y^1_{s(i)} ),$ 
as in (\ref{d(sigma)})
%Definition \ref{def:d}
.
%Since maps   agree on  on truncated edges, $d(  \Phi_0, \Phi_1, \sigma^{1\tau}_i)$ on
%truncated edges is  always zero.
It follows from Proposition \ref{prop:RelHomCell} that
\bge
\Phi_0\vert \bsigma^{1c}_i \simeq \Phi_1\vert \bsigma^{1c}_i \ rel. \ (\sigma^{1c}_i)^{\dd} \Leftrightarrow
d(  \Phi_0, \Phi_1,\sigma^{1c}_i,Y^1_{s(i)})=0.
\label{EdgeHomConvex}
\ene
%(compare with {\bf 4} and (8.1) in \cite{eilenberg}).
Define a formal linear combination
\bge
d^1_{} (\Phi_0,\Phi_1) = \dst\sum_{i} d(  \Phi_0, \Phi_1,\sigma^{1c}_i,Y^1_{s(i)} ) \sigma^{1c}_i.
\label{d1}
\ene
%as in (\ref{d}).
%(see Definition \ref{def:d}).
(The summation is over cleaved edges of $\Phat .$
Note that $\Phi_0 $ and $\Phi_1$ coincide  and are constant  on truncated edges, thus
$d(  \Phi_0, \Phi_1, \sigma^{1\tau}_i, Y )=0.$) It follows from Proposition \ref{prop:RelHomCell} that 
\bge
 \Phi_0\vert \Phat_1 \simeq \Phi_1\vert \Phat_1 \ rel. \ \Phat_0 \Leftrightarrow d^1_{} (\Phi_0,\Phi_1)=0.
 \ene
\noindent Thus if 
%$\Phi_0, \Phi_1:\Phat \rightarrow RP^2$ are tangent maps, and 
$d^1_{} (\Phi_0,\Phi_1)=0$, there exist tangent homotopy 
%\bge
$H_{\Phi_0,\Phi_1}^{1}: \Phat_1 \times E \rightarrow RP^2 ,$ such that  
%$H_{\Phi_0,\Phi_1}^{1}\vert  \Phat_1 \times \lbr m \rbr  = \Phi_m\vert \Phat_1  ,  m=0,1,$
%\label{H1}
%\ene

\bge
\bga
H_{\Phi_0,\Phi_1}^{1}\vert  \Phat_1 \times \lbr m \rbr  = \Phi_m\vert \Phat_1  ,  m=0,1,
\\
H_{\Phi_0,\Phi_1}^{1} \lb \bsigma^{1c}_i \times E\rb  \subset Y_{s(i)}\cong S^1 \ \mbox{ on cleaved edges,} 
\\
H_{\Phi_0,\Phi_1}^{1}\vert \lb \bsigma^{1t}_j \times E  \rb = \Phi_0\vert\bsigma^{1t}_j= \Phi_1\vert\bsigma^{1t}_j = q_j \ \mbox{on truncated edges,}
\label{H1}
\ena
\ene
where $q_j$ correspond to the constant director parallel to the edge  $\sigma^{1t}_j$. 
%!!! do we need it?
%We denote truncated edge homotopies by $H_{\Phi_0,\Phi_1}^{1\tau,i} = H_{\Phi_0,\Phi_1}^{1}\vert \bsigma^{1\tau}_i \times E$, and  cleaved edge homotopies by $H_{\Phi_0,\Phi_1}^{1c,i} = H_{\Phi_0,\Phi_1}^{1}\vert \bsigma^{1c}_i \times E$.
%!!! do we need it? Integral formulas?
%(On    truncated edges  $\sigma^{1\tau}_j,$ homotopy map is a constant map to a point on $RP^2$ corresponding to director parallel to  the truncated edge).
%with respect to homotopy parameter,  $H_{\Phi_0,\Phi_1}^{1,i}( \sigma^{1\tau}_i \times E)= \Phi_0\vert \sigma^{1\tau}_i = \Phi_1\vert \sigma^{1\tau}_i$ .)
Assume that \mbox{$d^1_{} (\Phi_0,\Phi_1)=0$.} For a face $\sigma^2 $  define a map $\Phi$ on  $(\sigma^2\times E)^{\dd} 
%\cong 
= S^2$
%(homeomorphic to $S^2$)
by

\bge
\bga
\Phi\vert (\sigma^2\times \{m\}) = \Phi_m, m= 0,1;
\\
\Phi\vert (\bsigma^1_i\times E) = H_{\Phi_0,\Phi_1}^{1}\vert (\bsigma^1_i \times E), \quad \bsigma^1_i \subset (\sigma^2)^{\dd} \quad .
%\\
%\Phi\vert\lb \lb\sigma^1_i\rb^{\dd} \times E\rb = \Phi_0\vert  \lb\sigma^1_i\rb^{\dd}   = \Phi_1\vert  \lb\sigma^1_i\rb^{\dd}  \quad .
\ena
\label{PConvexFaceHom}
\ene

%$H (\Phi_0,\Phi_1, \sigma^1_i \times E),$ on edges    $\sigma^1_i\subset \Phat$ so that  $H (\Phi_0,\Phi_1, \sigma^1_i \times 0)= \Phi_0\vert \sigma^1_i,$ $H (\Phi_0,\Phi_1, \sigma^1_i \times 1)= \Phi_1\vert \sigma^1_i$  (those homotopies are constant maps on truncated edges).  For a face $\sigma^2$ and $E=[0,1],$ define a map $\Phi$ on the sphere $(\sigma^2\times E)^{\dd}$  by
\begin{figure}[h]
\begin{center}
\input{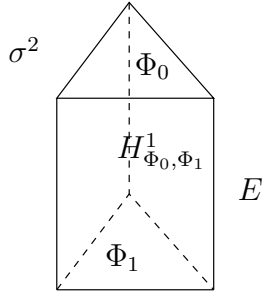}
\caption{Homotopy on faces.}
\label{fig:FaceHom}
\end{center}
\end{figure}

We can lift maps $\Phi_0$, $\Phi_1 $, and $\Phi$ to 
%continuous 
maps 
to $S^2$, $\myPhi_0$, $\myPhi_1$, $\myPhi$ . %the universal covering space for $RP^2$.
Such lifting is uniquely determined   by  selecting one of the two possible values for the lifting at a point $\tp$ on a truncated edge $\tsig$ of $\Phat$;  we choose  this  value to be the same for $\myPhi_0$ and  $\myPhi_1 $.
The lifted maps $\myPhi_0$ and  $\myPhi_1 $  will be constant  and equal on  $\tsig$ ,
\bge
\myPhi_0 (\tp)= \myPhi_1 (\tp) = \myPhi_0\vert \btsig = \myPhi_1\vert \btsig =   \mathbf{\hat{e}}_p,
\label{Phi0sPhi1s}
\ene
%where 
$\mathbf{\hat{e}}_p\in S^2$. 
%is 
%a unit vector 
%parallel to the  truncated edge $\tsig$.  
It then follows from (\ref{Phi0sPhi1s}) and $d^1_{} (\Phi_0,\Phi_1)=0$  that
\bge
\myPhi_0\vert \bsigma^{1t}_i  = \myPhi_1\vert \bsigma^{1t}_i =   \mathbf{\hat{e}}_i
\label{edgeOrient}
\ene
for {\em all}  truncated edges
%, not just the edge $\tsig$
.
(Unit vectors $\mathbf{\hat{e}}_i$ are parallel to respective $\bsigma^{1t}_i$ and determined up to an overall sign affecting all truncated edges, corresponding to two possible liftings 
%of   ${\Phi}_0$, ${\Phi}_1$  to     $\myPhi_0$, $\myPhi_1$ at $\tp$
in (\ref{Phi0sPhi1s}). 
$\mathbf{\hat{e}}_i$ are called edge orientations in \cite{rz}).
%If   $d^1_{} (\Phi_0,\Phi_1)=0, $
We lift  the map  $\Phi \vert (\sigma^2\times E)^{\dd}$ in (\ref{PConvexFaceHom})  to a map  to $S^2$, $\myPhi\vert (\sigma^2\times E)^{\dd}$  by requiring that
\bge
%\myPhi\vert \sigma^{1t}_i\times E= \myPhi_0\vert \sigma^{1t}_i = \myPhi_1\vert \sigma^{1t}_i
\myPhi (t_i \times E ) = \myPhi_0 (t_i) = \myPhi_1 (t_i) = \mathbf{\hat{e}}_i, \quad t_i \in \bsigma^{1t}_i \cap  \bsigma^2 .
\label{PhiS}
\ene
Such lifting defined for different faces is consistent on common edges, and defines the lifting of 
$H_{\Phi_0,\Phi_1}^{1}$  to a map $H_{\Phi_0,\Phi_1}^{ \uparrow 1}: \Phat_1 \times E\rightarrow S^2$, 
\bge
H_{\Phi_0,\Phi_1}^{ \uparrow 1}\vert (\bsigma^1_i \times E)=  \myPhi\vert (\bsigma^1_i\times E), \quad \bsigma^1_i \subset (\sigma^2)^{\dd} .
\label{H1S}
\ene
%This choice is consistent for all truncated and cleaved faces $\sigma^2$ and def

\begin{prop} Let $\Phi_0, \Phi_1:\Phat \rightarrow RP^2$ be tangent maps with  $d^1_{} (\Phi_0,\Phi_1)=0,$    $\Phi$ is given by (\ref{PConvexFaceHom}), and the lifted to  $S^2$ maps  $\myPhi_0$, $\myPhi_1 $,  $ \myPhi$  are  defined by  (\ref{Phi0sPhi1s})- (\ref{PhiS}).
\\
%Let $\Phi_0, \Phi_1: \Phat\rightarrow RP^2$ be tangent maps. $\Phi$ is given
1. Let $\sigma^{2c}$ be a { cleaved} face. 
%Let $c^3 (\myPhi,\sigma^{2c}\times E, S^2 )$ be an element of $\pi_2 (S^2),$ corresponding to the map $\myPhi \vert (\sigma^{2c}  \times E)^{\dd} \rightarrow S^2.$      Then
\\
a) %$c^3 (\Phi,\sigma^{2c}\times E , RP^2 )$
$c^3 ( \myPhi,\sigma^{2c}\times E, S^2 )$
does not depend from the choice of tangent homotopy  $H_{\Phi_0,\Phi_1}^{1}$ in (\ref{PConvexFaceHom}).
% on cleaved edges.
%on edges $\sigma^1_i \subset (\sigma^{2c})^{\dd}$.
\\
b) %$c^3 (\Phi,\sigma^{2c} \times E, RP^2  )=0$
$c^3 (\myPhi,\sigma^{2c}\times E, S^2 )=0$
 if and only if the map $\Phi$ can be extended to a map   \mbox{$ H_{\Phi_0,\Phi_1}^{2c}: \bsigma^{2c} \times E\rightarrow RP^2.$}   The latter is equivalent to  $\Phi_0\vert \bsigma^{2c}  \simeq \Phi_1\vert \bsigma^{2c} $ .
\\
2. Let  $\sigma^{2\tau}_i$ be  a { truncated face,} let $S^1_i$ be the great circle of $S^2$ parallel to  $\sigma^{2\tau}_i$, then
\bge
c^3 (\myPhi,\sigma^{2\tau}_i\times E, S^1_i ) \equiv 0 .
\label{c3tau0}
\ene
%an element of %$\pi_2 (Y^1_i),$ $\pi_2 (S^1),$ corresponding to the map  $\myPhi\vert (\sigma^{2\tau}_i \times E)^{\dd} \rightarrow S^1_i , $ is always zero as $\pi_2(S^1)=0$.
It follows that the  tangent homotopy $H_{\Phi_0,\Phi_1}^{1}$ 
%on  $\Phat_1$ 
%of a truncated face %$\sigma^1_i\subset \sigma^{2\tau}^{\dd} $
can always be extended to  homotopies on  truncated faces $H_{\Phi_0,\Phi_1}^{2\tau,i}: \bsigma^{2\tau}_i \times E \rightarrow Y^1_i $.
%$H_{\Phi_0,\Phi_1}^{2t,i} (\sigma^{2\tau}_i \times E)\rightarrow Y^1_i $ .
\label{Prop:SurfaceHomConvex}
\end{prop}
\noindent Proof.
%\\
%1. A map $S^2\rightarrow RP^2$ can be lifted to  a map to a covering space  $S^2\rightarrow S^2 $ (by assigning a unit vector to a director at one point arbitrary, and extending to other points by continuity). Such map is homotopic to a constant map if and only if the lifted map is homotopic to a constant map (see e.g. \cite{novikov}). A map $S^2\rightarrow S^2$ is classified by a degree.
%\\
Let $H_{\Phi_0,\Phi_1}^{1}$, $\tilde{H}_{\Phi_0,\Phi_1}^{1}$  be two tangent edge homotopies, which by construction satisfy $H_{\Phi_0,\Phi_1}^{1} \vert (\Phat_1\times E)^{\dd} = \tilde{H}_{\Phi_0,\Phi_1}^{1}\vert (\Phat_1\times E)^{\dd}$. Let $\tilde{\Phi}$  be given by (\ref{PConvexFaceHom}), with     $H_{\Phi_0,\Phi_1}^{1}$ replaced by $\tilde{H}_{\Phi_0,\Phi_1}^{1}$, and  
$\tilde\myPhi$, $\tilde{H}_{\Phi_0,\Phi_1}^{\uparrow 1}$ lifted to $S^2$ maps, 
$\tilde\myPhi\vert \sigma^{1t}_i\times E=  \tilde{H}_{\Phi_0,\Phi_1}^{\uparrow 1} \vert \sigma^{1t}_i\times E= \myPhi\vert \sigma^{1t}_i\times E = \myPhi_0\vert \sigma^{1t}_i = \myPhi_1\vert \sigma^{1t}_i$ on all truncated edges  $\sigma^{1t}_i$.
%Lifted maps  ${H}_{\Phi_0,\Phi_1}^{ \uparrow 1},$  $\tilde{H}_{\Phi_0,\Phi_1}^{\uparrow 1}$ are defined as in (\ref{H1S}).
\\
1a). By construction $H_{\Phi_0,\Phi_1}^{1}$, $\tilde{H}_{\Phi_0,\Phi_1}^{1}$ are constant, equal  maps on truncated edges. On cleaved edges, \\ $ 
{H}_{\Phi_0,\Phi_1}^{\uparrow 1 }  \vert \lb  \sigma^{1c}_i \times E \rb^{\dd} = \tilde{H}_{\Phi_0,\Phi_1}^{ \uparrow 1}\vert \lb  \sigma^{1c}_i \times E \rb^{\dd}$ .
%\  (indeed, both homotopies  equal $\myPhi_m$ on $\sigma^{1c}_i \times \lbr m \rbr$, $m=0,1$, and on  $\lb \sigma^{1c}_i \rb^{\dd}  \times E$  are constant, equal  maps to the edge orientation on corresponding truncated edge (\ref{edgeOrient})). 
It follows from (\ref{cISd}) and (\ref{d:addition}) that
\bge
c^3 ( \myPhi,\sigma^{2c}\times E , S^2 ) - c^3 ( \tilde\myPhi,\sigma^{2c}\times E , S^2 )= \dst\sum_{ i: \sigma^{1c}_i \subset \bsigma^{2c} } d \lb {H}_{\Phi_0,\Phi_1}^{\uparrow 1} , \tilde{H}_{\Phi_0,\Phi_1}^{ \uparrow 1},   \sigma^{1c}_i \times E, S^2\rb = 0 .
\label{cdHfaceEdge1}
\ene
The last equality is due to the fact that on  $\sigma^{1c}_i \times E$,  ${H}_{\Phi_0,\Phi_1}^{1} $, $\tilde{H}_{\Phi_0,\Phi_1}^{1,i} $ are maps to $S^1$, the great circle  of $S^2$   parallel to the truncated face containing $\sigma^{1c}_{i}$, and $\pi_2(S^1) =0 .$

%Projecting back to $RP^2$,
%Indeed, a map  $S^2\rightarrow RP^2$ can be lifted to  a map to a covering space  $S^2\rightarrow S^2 $ (by assigning a unit vector to a director at one point arbitrary, and extending to other points by continuity). Such map is homotopic to a constant map if and only if the lifted map is homotopic to a constant map (see e.g. \cite{novikov}). If we lift the map  $({H}_{\Phi_0,\Phi_1}^{1,i} ,  \tilde{H}_{\Phi_0,\Phi_1}^{1,i} , \sigma^{1c}_i \times E, RP^2)$ to a map $S^2\rightarrow S^2$, the image will be on the great  circle parallel to the truncated face of $\Phat$ containing  $\sigma^{1c}_i$ . Therefore, we can find a point $p$ on $S^2$ which is not in the image of the lifted map, and therefore the lifted map is homotopic to a constant map (a homotopy may be obtained by contracting the image to a point diametrically opposite to $p$ ).
% regular value.
%The value of the map on $\sigma^{1c}_i\times E,$ where $\sigma^{1c}_i$ are cleaved edges of the cleaved face $\sigma^{2c},$
%$\sigma^1_i \subset (\sigma^2)^{\dd}$
%is always in the union of great circles $\bigcup S^1_i,$ parallel to truncated faces intersecting $\sigma^{2c}.$  One can always choose a regular value  not in $\bigcup S^1_i,$ therefore degree of the lifted map is not affected by a choice of tangent homotopy on edges $H_{\Phi_0,\Phi_1}^{1,i} (\sigma^1_i \times E).$

\noindent 1b). $(\sigma^{2c} \times E)^{\dd}$
%is homeomorphic to $S^2$,
$ \cong S^2$,
and if the element of $\pi_2 (S^2)$ corresponding to   $\myPhi: (\sigma^{2c} \times E)^{\dd} \rightarrow S^2$ is zero, then $\myPhi$ can be extended to a map of $\bsigma^{2c} \times E \cong D^3$
%(homeomorphic to a ball $B^3$)
to $S^2$. Such an extension
%composed with
followed by the projection from $S^2$ to $RP^2$  provides the required homotopy.
\\
2. $\pi_2 (S^1)=0,$ thus a map
%$\Phi: S^2\rightarrow S^1$
$\myPhi\vert (\sigma^{2\tau}_i \times E)^{\dd} \rightarrow S^1_i $
can be always extended to a map $\bsigma^{2\tau}_i \times E \rightarrow S^1_i.$ Such an extension
followed by the projection  to $Y_1^i\subset RP^2$
provides the required homotopy.
\qed

%If $d^1_{} (\Phi_0,\Phi_1) = 0$, we can define  $c^3 (\myPhi,\sigma^{2c}\times E, S^2 )$ on cleaved faces by (\ref{PConvexFaceHom})-(\ref{PhiS}). The map $\myPhi$  is defined via  maps $\Phi_0, \Phi_1$, tangent edge homotopy $H^{1}_{\Phi_0, \Phi_1}$, and a choice of  lifting (\ref{Phi0sPhi1s})-(\ref{PhiS}), however it follows from Proposition \ref{Prop:SurfaceHomConvex}   that different choices of $H^{1}_{\Phi_0, \Phi_1}$ yields the same $c^3 (\myPhi,\sigma^{2c}\times E, S^2 )$.

\begin{defn} Assume that $d^1_{} (\Phi_0,\Phi_1) = 0$. We define $d^2_{} (\Phi_0,\Phi_1)$ by
\bge
d^2_{} (\Phi_0,\Phi_1) =   \dst\sum_i c^3 (\myPhi,\sigma^{2c}_i \times E, S^2 ) \sigma^{2c}_i \quad ,
\label{d2}
\ene
where 
%summation is over cleaved faces of ${\Phat}$, and  
$c^3 (\myPhi,\sigma^{2c}\times E, S^2 )$ is defined on cleaved faces by (\ref{PConvexFaceHom})-(\ref{PhiS}). $d^2_{} (\Phi_0,\Phi_1)$ does not depend from the choice of edge homotopy $H^{1}_{\Phi_0, \Phi_1}$, and is defined up to an overall sign, as there are two ways of lifting in (\ref{Phi0sPhi1s})-(\ref{PhiS}).
%(\ref{Phi0sPhi1s}).
\label{Def:d2}
\end{defn}
\noindent If $d^1 (\Phi_0,\Phi_1)=0$ and $d^2 (\Phi_0,\Phi_1) = 0$, by Proposition \ref{Prop:SurfaceHomConvex} there exist a tangent homotopy 
\bge
H_{\Phi_0,\Phi_1}^{2}: \Phat_2 \times E \rightarrow RP^2,  H_{\Phi_0,\Phi_1}^{2}\vert  \Phat_2 \times \lbr m \rbr  = \Phi_m\vert \Phat_2  ,  m=0,1,
\label{H2}
\ene
given  by $H_{\Phi_0,\Phi_1}^{2}\vert\bsigma^{2a} \times E = H_{\Phi_0,\Phi_1}^{2a}$, $a=c,t$ ($H_{\Phi_0,\Phi_1}^{2a}$ were introduced in Proposition \ref{Prop:SurfaceHomConvex}). Moreover,  by construction 
\bge
H_{\Phi_0,\Phi_1}^{2}\vert\Phat_1 \times E =  H_{\Phi_0,\Phi_1}^{1} \quad .
\label{H2H1}
\ene  

\begin{rem}
\noindent
%$c^3 (\myPhi,\sigma^{2c}\times E, S^2)=0$
In terms of invariants in \cite{rz}, $d^1_{} (\Phi_0,\Phi_1)=0$ means that   {\em kink numbers} of $\Phi_0, \Phi_1$ are the same, and {\em edge orientations} are the same up to a  simultaneous  change of sign on all edges. $d^2_{} (\Phi_0,\Phi_1) = 0$
means that $\Phi_0,\Phi_1$ have the same {\em wrapping numbers} on cleaved faces.
%$\sigma^{2c},$  in terms of invariants in \cite{rz}.
\end{rem}
\begin{figure}[h]
\begin{center}
\input{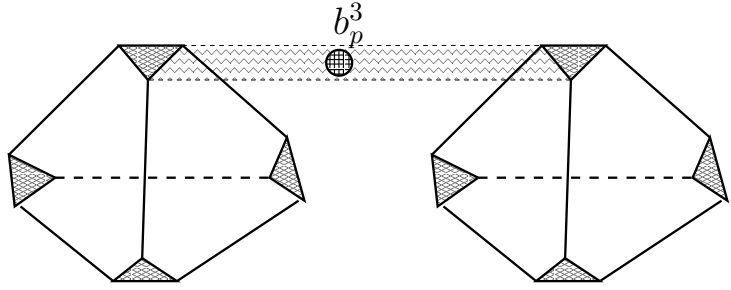}
\caption{Bulk homotopy: no Hopf invariant (escape through a vertex).}
\label{fig:BulkHom}
\end{center}
\end{figure}
\begin{thm}
%\noindent {\bf \thm (Homotopy classification).}
Tangent maps $\Phi_0, \Phi_1$ are homotopic if and only if
%$d^1_{} (\Phi_0,\Phi_1)=0$ and \mbox{$d^2 (  \Phi_0, \Phi_1 )=0$,} %which is equivalent to
%for every truncated face $\sigma^{2t}_j$, for all cleaved edges    $\sigma^{1c}_i\subset \lb \sigma^{2t}_j\rb^{\dd}$, we have that
\bge
d^1_{} (\Phi_0,\Phi_1)=0,
%%d (  \Phi_0, \Phi_1, \sigma^{1c}_i, Y^1_j )=0;
\label{kinkP}
\ene
%and %for all cleaved  faces  $\sigma^{2c}_i$ we have that
\bge
%%c^3 (\myPhi,\sigma^{2c}_i\times E , S^2 )=0.
d^2_{} (\Phi_0,\Phi_1) = 0.
\label{wrapP}
\ene
%(Here $d^1_{} (\Phi_0,\Phi_1)$ is defined by (\ref{d1}), and $d^2_{} (  \Phi_0, \Phi_1 )$ is defined by  (\ref{d2})).
%the map $\Phi$ is defined in (\ref{PConvexFaceHom}), and $\myPhi$ is   lifting of $\Phi$ to a map to $S^2$).
\label{ThmConvexHom}
\end{thm}
\noindent {\em Proof.} It is
%obvious
clear that (\ref{kinkP}), (\ref{wrapP})  are necessary.
To show they are sufficient, note that from Proposition \ref{Prop:SurfaceHomConvex}, there exist face homotopy $H_{\Phi_0,\Phi_1}^{2}$ (\ref{H2}). %\bsigma^{2c}_i\times E \rightarrow RP^2$ on cleaved faces $\sigma^{2c}_i,$ tangent on cleaved edges, and tangent homotopies $H_{\Phi_0,\Phi_1}^{2t,i}: \bsigma^{2\tau}_i\times E \rightarrow Y^1_i$ on truncated faces $\sigma^{2\tau}_i $. (Those surface homotopies by construction agree on  all cleaved and  truncated edges).
Define a map $\Phi$ on
$(\Phat\times E)^{\dd} \cong S^3$ by
\bge
\bga
\Phi\vert (\Phat\times \{m\}) = \Phi_m, m= 0,1,
\\
\Phi\vert (\bsigma^{2a}_i\times E) = H_{\Phi_0,\Phi_1}^{2}\vert \bsigma^{2a}_i \times E \equiv H_{\Phi_0,\Phi_1}^{2a,i} ,  
%= H_{\Phi_0,\Phi_1}^{2a,i}, 
\quad \bsigma^{2a}_i \subset (\Phat)^{\dd}  
% a= c,t
.
\ena
\label{PhiWrapNum}
\ene
%(where $H$ are surface homotopies. )
Let $c^4\equiv c^4 (\Phi,\Phat\times E, RP^2 )$ be the corresponding element of $\pi_3 (RP^2)= \Bbb Z$.  %(as $RP^2$ is 3-simple, and we need not be concerned about the base point). 
If $c^4=0$, $\Phi$ can be extended to a map of  $(\Phat\times E)$ to $RP^2$, and thus $\Phi_0$ and  $\Phi_1$ are homotopic.  $c^4 $ depends from the maps $\Phi_0,$ $\Phi_1,$ and face homotopy $H_{\Phi_0,\Phi_1}^{2}$ and in general need not to be zero.  However, we can  always modify $H_{\Phi_0,\Phi_1}^{2c}$ on $\sigma^{2c}\times E_0$ for a selected cleaved face $\sigma^{2c}$, so that $c^4$ becomes zero. This process may be described briefly as gluing an $S^3$ to a point   $p\in \sigma^{2c}\times \eo $, and taking a map on   $S^3$   such that the  corresponding element of $\pi_3 (RP^2)$ equals
$-c^4 $. In more detail, we first make some room for such modification, by taking a point   $p \in \sigma^{2c}\times \eo$ and expanding  it to a ball $b_p^3\subset  \sigma^{2c}\times \eo$. It is clear that there is a map
%$H_{\Phi_0,\Phi_1}^{2c}$ to a  new map
$\tilde{H}_{\Phi_0,\Phi_1}^{2c}$,  homotopic to $H_{\Phi_0,\Phi_1}^{2c}$, constant on $b_p^3$, 
$\tilde{H}_{\Phi_0,\Phi_1}^{2c}\vert b_p^3  \equiv   q = H_{\Phi_0,\Phi_1}^{2c}(p)$ and   unchanged on the boundary,  $\tilde{H}_{\Phi_0,\Phi_1}^{2c}\vert(\sigma^{2c} \times E)^{\dd}=  {H}_{\Phi_0,\Phi_1}^{2c}\vert(\sigma^{2c} \times E)^{\dd}$. (Indeed, up to a homeomorphism we can take $\bsigma^{2c}\times E$ to be a 3-ball $B^3$  centered at  $p$ and $b_p^3$  a ball of smaller radius with the same center; then on the spherical shell $B_p^3\setminus b_p^3$  we take $\tilde{H}_{\Phi_0,\Phi_1}^{2c}$ to be a radially rescaled  map ${H}_{\Phi_0,\Phi_1}^{2c}$).
%It is clear that we can have  $\tilde{H}_{\Phi_0,\Phi_1}^{2c}$    homotopic to ${H}_{\Phi_0,\Phi_1}^{2c}$ (indeed, $\sigma^{2c}\times E$ is homeomorphic to a 3-ball, and the region $\sigma^{2c}\times E\setminus b_p^3$   to a spherical shell;
%Indeed,
%assume that $c^4 (\Phi,\Phat\times E )= n,$ and $\Phi\vert (\sigma^{2c}\times E) = H_{\Phi_0,\Phi_1}^{2c} %(\sigma^{2c} \times E).$
%Let $b_p^3 \subset\subset \sigma^{2c}\times E$ be a 3-ball with center at a point $p\in \sigma^{2c}\times E$ and strictly in the interior of $\sigma^{2c}\times E.$ There is a map $\tilde{H}_{\Phi_0,\Phi_1}^{2c},$ which coincide with ${H}_{\Phi_0,\Phi_1}^{2c}$ on $(\sigma^{2c} \times E)^{\dd},$ is homotopic to ${H}_{\Phi_0,\Phi_1}^{2c},$ and is a constant map to  $q={H}_{\Phi_0,\Phi_1}^{2c} (p)$ on the ball $b_p^3 .$ ( $\tilde{H}_{\Phi_0,\Phi_1}^{2c}$    outside of $b_p^3$ is a shrinked  in the radial direction version of ${H}_{\Phi_0,\Phi_1}^{2c}.$ )
We now modify  $\tilde{H}_{\Phi_0,\Phi_1}^{2c}$ on $b_p^3$: let  $\varphi_{-n} $ be a map $b_p^3\rightarrow RP^2,$ with $(b_p^3)^{\dd} \mapsto q\in RP^2$, such that the element of $\pi_3 (RP^2)$ corresponding to this map is $-n = - c^4 .$ Define $\tilde{\Phi}$ by
\bge
\bga
\tilde{\Phi}\vert \lb ( \bsigma^{2c} \times E ) \setminus  b_p^3 \rb  = \tilde{H}_{\Phi_0,\Phi_1}^{2c},
\\[1mm]
\tilde{\Phi}\vert  b_p^3  =  \varphi_{-n},
\\[1mm]
\tilde{\Phi}\vert \lb(\Phat\times E)^{\dd} \setminus  \lb \sigma^{2c} \times \eo \rb \rb  = \Phi .
\label{glue}
\ena
\ene
Then $\tilde{c}^4 \equiv c^4 (\tilde{\Phi},\Phat\times E, RP^2 )= 0,$ and so $\tilde{\Phi}$ can be extended to a map $\Phat\times E\rightarrow RP^2,$ providing a tangent homotopy between maps $\Phi_0$ and $\Phi_1.$
\qed

%Let's modify $\Phi$ as fo $\tilde{H}_{\Phi_0,\Phi_1}^{2c}$  on $b_p^3:$ we define a new map which is $\tilde{H}_{\Phi_0,\Phi_1}^{2c}$

%(b_p^3,(b_p^3)^{\dd}, RP^2,q)

%modify $\Phi\vert (\Phat\times \{0\})$ on a small ball $B^3\subset \Phat,$ with $\Phi\vert \partial B^3$ being a constant, so that for the modified map $\tilde{\Phi}$ we have that  $c^4 (\tilde{\Phi},\Phat\times E )=0, $ and therefore $\tilde{\Phi}$ can be extended to $ \Phat\times E.$ This means that
%$\Phi_1$ is homotopic to $\tilde{\Phi}_0,$ where $\tilde{\Phi}_0 $  is obtained from ${\Phi}_0$ by modifying on a small ball   as above. That ball then can be removed from $\Phat$ through a cleaved face  by a homotopy, as in \cite{rz}.
%\vspace{40mm}
Condition $d^1_{} (\Phi_0,\Phi_1)=0 $ in (\ref{kinkP}) is equivalent to $d(  \Phi_0, \Phi_1,\sigma^{1c}_i,Y^1 ) =0$ for all cleaved edges $\sigma^{1c}_i$. Those conditions are not independent, there is one relation (\ref{kinkSum}) among them per every truncated face. Similarly, condition $d^2(  \Phi_0, \Phi_1)=0 $ in   (\ref{wrapP}) is equivalent to $d(  \myPhi_0, \myPhi_1,\sigma^{2c}_i, S^2) =0$ for all cleaved faces $\sigma^{2c}_i$; there is one overall relation (\ref{wrapSum}) between those conditions .

\begin{prop}{\bf (Sum rules).} \\{\em
1.
Let $\sigma^{2\tau}_i$ be a truncated face. Orient  cleaved edges $\sigma^{1c}_j$   of $\sigma^{2\tau}_i $ consistent with $\partial \sigma^{2\tau}_i.$ Let $\Phi_0$ and $\Phi_1$ be two tangent maps. Since restrictions of  $\Phi_0$  and  $\Phi_1$ to $\sigma^{2\tau}_i$ are continuous, tangent  maps to $Y^1_i$, we have that
\bge
c^2 (\Phi_m, \sigma^{2\tau}_i, Y^1_i) = 0, m= 0,1.
\label{kinkSR}
\ene
%We note that (\ref{kinkSR}) is equivalent to sum rules for kink numbers of maps $\Phi_0$  and  $\Phi_1$, as introduced in \cite{rz}.
Using addition formula (\ref{d:addition}), we get
\bge
 \dst\sum_{j:\sigma^{1c}_j\in (\sigma^{2\tau}_i)^{\dd} } d(  \Phi_0, \Phi_1, \sigma^{1c}_j,Y^1_i ) =  c^2 (\Phi_0, \sigma^{2\tau}_i, Y^1_i)-c^2 (\Phi_1, \sigma^{2\tau}_i, Y^1_i)=0.
\label{kinkSum}
\ene
%This identity implies that for every truncated face, the number of independent conditions  in (\ref{kinkP}) is one less then the number of cleaved edges on that face.
\\
2. Continuity of  $\Phi_0$  and  $\Phi_1$ on $\Phat$ implies
\bge
c^3 (\myPhi_m,\Phat , S^2) = 0, m= 0,1,
\label{wrapSR}
\ene
where $\myPhi_m$ is lifted to $S^2$ map.
%, as in  (\ref{PConvexFaceHom}), (\ref{PhiS}).
%which is equivalent to sum rule for wrapping numbers \cite{rz}.
Assume that  $d^1_{} (\Phi_0,\Phi_1)=0 $.
Orient cleaved faces $\{\sigma^{2c}_i\}$ consistent with $\partial \Phat.$ Then using (\ref{cISd}), (\ref{d:addition}), (\ref{c3tau0}) we get
\bge
\dst\sum_{\sigma^{2}_i\in (\Phat)^{\dd} } c^3 (\myPhi,\sigma^{2}_i\times E , S^2 )= \dst\sum_{\sigma^{2c}_i\in (\Phat)^{\dd} } c^3 (\myPhi,\sigma^{2c}_i\times E , S^2 )=  c^3 (\myPhi_0,\Phat , S^2)  - c^3 (\myPhi_1,\Phat , S^2) =0.
\label{wrapSum}
\ene
%(here the map $\Phi$ is defined in (\ref{PConvexFaceHom})).
%This identity implies that  the number of independent conditions  in (\ref{wrapP})   is one less then the number of cleaved  faces.
}
\label{Prop:SR}
\end{prop}
%This follows from addition rules (see (8.2) in \cite{eilenberg}) and the fact that  the maps are continuous on truncated faces and on $\Phat$. Another proof is given in \cite{rz}.
\section{Topological classification of post-aligned   nematic}
\begin{figure}[h]
%\begin{center}
%\includegraphics[natwidth=3in, natheight=2.in]{posttop.bmp}
\input{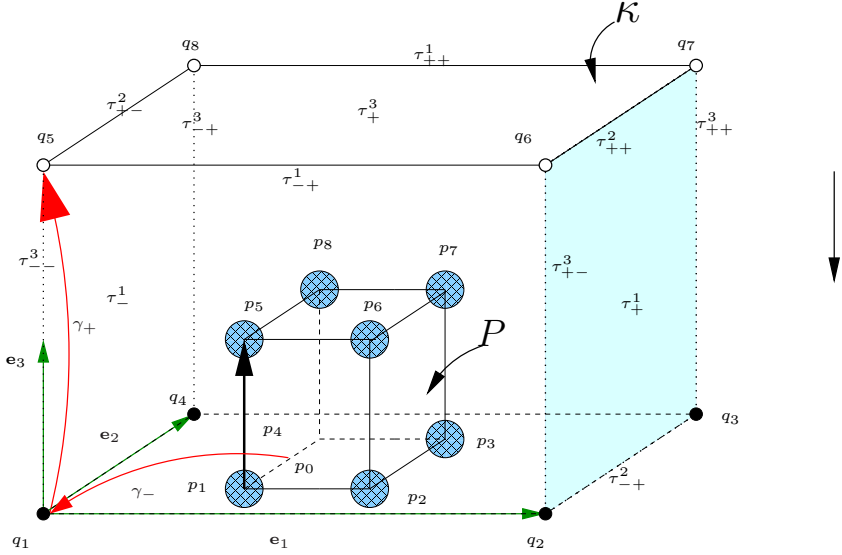}
\caption{Post geometry (fundamental domain with respect to horizontal translations). }
\label{figPost}
%\end{center}
\end{figure}

We now consider topological classification of a nematic liquid crystal in a domain $D$  between two horizontal plates and outside  a periodic array of rectangular posts situated on the bottom plate, Figure \ref{figPost}. Boundary conditions are tangent on the bottom plate and the post surface, and  periodic with respect to integer translations in horizontal X,Y directions. We consider normal (N) or tangent (T) boundary conditions on the top plate. Thus we consider homotopy classification of maps $D\rightarrow RP^2$, i.e.  director fields, satisfying the above boundary conditions, and continuous, except at the vertices of rectangular post (where boundary conditions disallow continuity). 

Let $\qq$ be the 
%closure of the 
fundamental domain of $D$ with respect to horizontal translations, i.e. a part of $D$ inside a rectangular prism with unit length and width, and  $\qqh$ be its truncated version, $\qqh = 
%\overline
{{\qq}\setminus \cup_i b_i}$, where $b_i= B_{i,\epsilon_i}\cap \qq$, and $B_{i,\epsilon_i}$ are small enough open 3-balls 
centered at the vertices of the rectangular post.
%  as in Section 2. 
As in Proposition 2.1, two maps $\qq\rightarrow RP^2$ are homotopic if and only if their restrictions to  the truncated region $\qqh$ are homotopic. 

It is clear that a periodic in $X,Y$ map $\Phi:\qqh\rightarrow RP^2$ can be lifted to a map $\myPhi: \qqh\rightarrow S^2,$ which is periodic or anti-periodic with respect to $X$, $Y$ translations. Indeed, extend the map a periodic map of the universal covering space of  $\qqh$ (the truncated version of $D$), lift it to a map to $S^2$, and then restrict to $\qqh$. Since the composition  of the resulting map with the projection to $RP^2$,  identifying diametrically opposite points of $S^2$,  is periodic,  $\myPhi$ is either periodic or anti-periodic.

Let $\Phi_0$, $\Phi_1$ be two maps $\qqh \rightarrow RP^2$ satisfying boundary conditions. It is clear that for $\Phi_0$ and $\Phi_1$ to be homotopic, we must have 
\bge
\dpst^1  (\Phi_0,\Phi_1)=0, \ \dpst^2 (\Phi_0,\Phi_1)=0,
\label{d=0Post}
\ene 
where $\dpst^1  (\Phi_0,\Phi_1),$ $\dpst^2  (\Phi_0,\Phi_1)$ are invariants associated with the vertices of the post and constructed as in Section 3. %Other invariants are associated with the boundary of $\qq$, as we explain below. 

Let $\kk$ denote the surface of the prism  bounding $\qqh .$ We can assume it is a unit prism with the center at the origin. We label vertices of $\kk$ as  $\ct^0_{\xi_1 \xi_2 \xi_3}= (\xi_1 \half , \xi_2 \half , \xi_3 \half) $, $\xi_i= \pm$ . Here and below let $\al, \be,\ga$ be a cyclic  permutation of $1,2,3$. 
%Let $\al, \be, \ga$ denote a cyclic permutation of $x,y,z$. 
We label an edge $x_\be = \xi_\be \half  , x_\ga = \xi_\ga \half $ as $\ct^1_{\al \xi_\be \xi_\ga}$, $\xi_\be, \xi_\ga = \pm$. We label a face $x_\al = \xi_\al \half $ as $\ct^2_{\al \xi_\al}$. Our maps are not defined on the footprint of the truncated post on the bottom face of $\kk$; this however will not play any role. In fact, due to the periodic boundary conditions and the addition formula (\ref{cISd}), (\ref{d:addition})   we have that 
$ c^2(\Phi_0, \ct^2_{3  -}, S^1)= c^2(\Phi_1, \ct^2_{3  -}, S^1)=0$, thus restrictions of maps $\Phi_0$, $\Phi_1$ to the bottom plate have tangent extensions from the boundary of the truncated post footprint to its interior. 

Select a point $p_0$ on a truncated edge, e.g. $p_0\in (p_1,p_4)$ (Figure \ref{figPost}). Let $\gamma_-\subset \qqh $ be a path on the bottom plate  from $p_0$ to the vertex $q_1 \equiv \ct^0_{---}$, which does not intersect any cleaved regions, and let $\gamma_+$ be the path from $q_1$ to $q_5$ along the vertical edge $\ct^1_{3--}$. Let $H^0_-$, $H^0_+$  be  paths  $E\rightarrow Y^1_\shortparallel \cong S^1$ in the space of directors parallel to horizontal plane such that
\bge
H^0_- (m) =  \Phi_m (q_1) = \Phi_m ( \ct^0_{\xi_1 \xi_2 -}) , \quad H^0_+ (m) = \Phi_m (q_5)= \Phi_m ( \ct^0_{\xi_1 \xi_2 +}) , m=0,1.
\label{H0End}
\ene
Define $\Phi :$ \  $( \gamma_- \times E)^{\dd}\cong S^1 \rightarrow Y^1_\shortparallel \cong S^1$ ,  $( \gamma_+ \times E)^{\dd}\cong S^1 \rightarrow RP^2$ by
\bge
\bga
\Phi\vert (  \gamma_\pm\times \{m\})  = \Phi_m\vert\gamma_\pm , m=0,1;
\\
\Phi\vert ( p_0 \times E) = \Phi_0 (p_0) =  \Phi_1 (p_0) ,
\\
\Phi\vert ( q_1 \times E) = H^0_- (E), \quad \Phi\vert ( q_5 \times E) = H^0_+ (E).
\ena
\label{HTgamma}
\ene
%Moreover,  select path  $H^0_-$ such that the elements of $\pi_1 (S^1)$ corresponding to \linebreak $\Phi\vert ( \gamma_- \times E)^{\dd}$ is trivial, and path $H^0_+$ such that the elements of $\pi_1 (RP^2)$ corresponding to \linebreak $\Phi\vert ( \gamma_+ \times E)^{\dd}$ is trivial

\noindent Moreover, select paths  $H^0_-$, $H^0_+$  such that the elements  of $\pi_1 (S^1)$ corresponding to  \linebreak $\Phi\vert ( \gamma_- \times E)^{\dd}$,  and    of $\pi_1 (RP^2)$ corresponding to   $\Phi\vert ( \gamma_+ \times E)^{\dd}$ are both trivial,
\bge
c^2_{\ga_-} (\Phi_0, \Phi_1) \equiv c^2(\Phi, \gamma_- \times E, S^1) =0, \quad c^2_{\ga_+} (\Phi_0, \Phi_1) \equiv c^2(\Phi, \gamma_+ \times E, RP^2) =0.
\label{H0trivial}
\ene
Define  map $(\Phi_0,\Phi_1): (\ct^1_{\al \xi_\be \xi_\ga} \times E)^{\dd} \cong  S^1 \rightarrow S^1,$ $\al =1,2$ 
%and  $(\ct^1_{3} \times E)^{\dd} \cong  S^1 \rightarrow RP^2,$ 
by
\bge
\bga
(\Phi_0,\Phi_1)\vert \lb \ct^1_{\al \xi_\be \xi_\ga} \times \{m\}\rb = \Phi_m\vert\ct^1_{\al \xi_\be \xi_\ga}, \ \al =1,2,   
%\quad (\Phi_0,\Phi_1)\vert \lb \ct^1_3 \times \{m\}\rb = \Phi_m\vert\ct^1_3 
\   m=0,1;
\\
(\Phi_0,\Phi_1) \vert \lb  \ct^0_{\xi_x \xi_y \pm}    \times E \rb = H^0_\pm (E) \quad .
\label{HTsigma1}
\ena
\ene
%%%%%%%end hom vertices
%%%%%%%hom edges
Let $d(  \Phi_0, \Phi_1,\ct^1_{\al \xi_\be \xi_\ga} ) $ , $\al=1,2$ be the corresponding  elements of $\pi_1 (S^1) $. In fact due to periodic boundary conditions it is enough to consider only edges  not related by horizontal translations. In the case of (T) boundary conditions on the top plate  we define $\dk(  \Phi_0, \Phi_1)$ as 
%$\ct^1_{\al \pm}$, where 
%and $d(  \Phi_0, \Phi_1, \ct^1_{3}) $ be the element of $\pi_1 (RP^2)$ 
%to those maps.
%non standard. Standard would be c^2 ( (\Phi_0,\Phi_1), \ct^1_\al\times E, RP^2) 
  
\bge
\dk (  \Phi_0, \Phi_1) = \dst\sum_{\al=1}^2 \dst\sum_{\sigma = +,-} d(  \Phi_0, \Phi_1, \ct^1_{\al \sigma} ) \ct^1_{\al \sigma},  \quad \  \ct^1_{1 \pm} \equiv \ct^1_{1 - \pm}, \ct^1_{2 \pm} \equiv \ct^1_{2  \pm -} ,  \quad (T) .
\label{d1kT}
\ene
Note that $(\Phi_0,\Phi_1)\vert(\ct^1_{\al +} \times E)^{\dd}$ and $(\Phi_0,\Phi_1)\vert(\ct^1_{\al -} \times E)^{\dd}$ are free-homotopic as $S^1\rightarrow RP^2$ maps, thus 
$d(  \Phi_0, \Phi_1, \ct^1_{\al +})  = d(  \Phi_0, \Phi_1, \ct^1_{\al -}) \ mod\  2$. In the case of (N) boundary conditions on the top plate $d(  \Phi_0, \Phi_1, \ct^1_{\al +} ) =0, \al =1,2 $ , thus
\bge
\dk (  \Phi_0, \Phi_1) = \dst\sum_{\al=1}^2  d(  \Phi_0, \Phi_1, \ct^1_{\al -} ) \ct^1_{\al -} .  \qquad  \qquad \qquad  \qquad \qquad \qquad\qquad (N) .
\label{d1kN}
\ene
Assuming (\ref{H0trivial}), it  is clear  that
\bge
 \Phi_0\vert \kk_1 \simeq \Phi_1\vert \kk_1  \Leftrightarrow d^1_{\kk} (  \Phi_0, \Phi_1)=0
\ene
(as usual, $\kk_n$ is the subcomplex of  $\kk$  consisting of cells of   $\kk$ of dimension $\leq n$).
Thus if 
\bge
\dk (  \Phi_0, \Phi_1)=0 , 
\label{d1T}
\ene
%and assuming (\ref{H0trivial}), 
there exists tangent homotopy  
$\hk $  periodic with respect to horizontal translations:
%:  {\kk_1 \times E}\rightarrow  RP^2$ 
%such that
\bge
\bga
\hk :{\kk_1 \times E}\rightarrow  RP^2, \hk \vert (\ct^1_{\al \xi_\be \xi_\ga} \times E)^{\dd} = \lb \Phi_0, \Phi_1 \rb 
%\vert (\ct^1_{\al \xi_\be \xi_\ga} \times E)^{\dd} \  
,  \ 
\hk \vert (\ct^1_{\al \xi_\be \xi_\ga} \times E) \subset Y^1_\shortparallel \cong S^1 ,\al =1,2 .
\ena
\label{H1al}
\ene
Assume that (\ref{d=0Post}), (\ref{d1T}) are satisfied.
Let $\myPhi_0$, $\myPhi_1 , \hkup $ be lifted to  $S^2$ maps, periodic or antiperiodic with respect to horizontal translations and defined similarly to (\ref{Phi0sPhi1s}) -(\ref{PhiS}),
\bge
\myPhi_0 (p_0)= \myPhi_1 (p_0) =   \mathbf{\hat{e}}_{\bf 2},  \quad \hkup \vert{\ct^1_{\al \xi_\be \xi_\ga} \times \lbr m \rbr}
= \myPhi_m\vert  {\ct^1_{\al \xi_\be \xi_\ga} } \ m=0,1 .
\label{Phi0sPhi1sPer}
\ene
%Let $\tau^2_\al$, $\al =1,2$ be  faces of the prism bounding $\qqh$ and orthogonal to $X$, $Y$ axes,  $\tau^2_1= (q_1,q_5, q_8, q_4)$ , $\tau^2_2= (q_1,q_2, q_6, q_5)$, and let $\tau^2_{3}$ be the top face of that prism. 
Define $\myPhi\vert\lb\ct^2_{\al,  \varsigma_\al} \times E\rb^{\dd}$ by 
%, $\tau^2_2$ be the face $(q_1,q_5, q_8, q_4)$Define   $\myPhi \vert\lb \ct^2_{\al \varsigma_\al} \times E\rb^{\dd}$ by
\bge\bga
\myPhi \vert \lb \ct^2_{\al \xi_\al}  \times \{m\}\rb = \myPhi_m \vert \ct^2_{\al \xi_\al} , m=0,1 , 
%\mu =1,2,3
\\
\myPhi \vert\lb  \ct^1_{\be \varsigma_\ga \varsigma_\al} \times E \rb= \hkup \vert \lb  \ct^1_{\be \varsigma_\ga \varsigma_\al} \times E \rb
, \quad
%\quad \ct^1_{\be \varsigma_\ga \varsigma_\al} \subset (\ct^2_{\al \varsigma_\al})^{\dd} 
\myPhi \vert\lb  \ct^0_{\xi_1 \xi_2 \xi_3} \times E \rb= \hkup \vert \lb  \ct^0_{\xi_1 \xi_2 \xi_3} \times E \rb .
%\vert \lb \ct^1_\be\times E \rb
\label{Phi2EDotBox}
\ena\ene

\begin{lem} %Let  $H^0$ be chosen so that $c^2_\ga(\Phi_0, \Phi_1) =0$ (\ref{H0trivial}). 
Assume  $\dk (  \Phi_0, \Phi_1)=0$ . 
Let $\hk$ and $\hkt$ be two 1-cell homotopies as in (\ref{H1al}), $\hkup$ and $\hkupt$ their lifted  to $S^2$ versions (\ref{Phi0sPhi1sPer}), and $\myPhi$, $\myPhit$ the corresponding maps $\lb\ct^2_{\al,  \varsigma_\al} \times E\rb^{\dd} \rightarrow S^2$ defined by (\ref{Phi2EDotBox}). Let  
$d^1_3  = d (\hkt, \hk, \tau^1_{3 -- }\times E, S^2 )\in \Bbb Z$. Let $s_a$ be an element of $\pi_1 (RP^2)$ 
%(or equivalently $\pi_1 (S^1) \ mod\ 2 $) 
corresponding to $\Phi_0\vert \ct^1_{a \pm } \simeq \Phi_1\vert \ct^1_{a \pm}$ , $a =1,2$. Then 
\bge
\bga
c^3(\myPhit, \ct^2_{\al -} \times E, S^2)- c^3(\myPhi, \ct^2_{\al  -} \times E, S^2)=  \sum_{\be=1}^2 \eps_{\al \be} (1-(-1)^{s_\be}) d^1_3  ,\quad \al  =1,2.
\ena
\label{additionT2}
\ene
\label{LemHTsigma2}
\end{lem}

\vspace{-7mm} 

\noindent{\em Proof.}
%We lift maps $\Phi_0$, $\Phi_1$ , $H^{1\ga}$ to maps $\myPhi_0$, $\myPhi_1$ , ${H}^{\uparrow 1\ga} $ to $S^2$, as in (\ref{Phi0sPhi1sPer}).
%%-(\ref{H11als})
%Homotopy $H^{1\ga}$    is fixed on $(\ct^1_\ga\times E)^{\dd}$ by (\ref{HTsigma1}), so if $\tilde{H}^{1\ga}$ is a different homotopy, we have that 
It follows from  (\ref{HTsigma1}), (\ref{H1al}) that 
%$\hk\vert (\ct^1_a\times E)^{\dd} = \hkt\vert(\ct^1_a\times E)^{\dd}, a= x,y,z$. 
$\hk\vert (\kk_1\times E)^{\dd} = \hkt\vert(\kk_1\times E)^{\dd} $. 
For the lifted maps periodic/antiperiodic conditions imply
\bge
\bga
%\hkup \vert (\ct^1_{\al +}\times E) =
\hkup \vert (\ct^1_{3 \xi_1 \xi_2}\times E) =
(-1)^{\mathop{\sum}_{\al=1}^{2} \half (1+\xi_\al)s_\al } \hkup  \vert (\ct^1_{3--}\times E) , 
%\hkupt \vert (\ct^1_{\al +}\times E) =(-1)^{s_\be} \hkupt  \vert (\ct^1_{\al  -}\times E) 
\label{perAper}
\ena
\ene
and the same condition for $\hkupt$. Using (\ref{cISd}), (\ref{Phi2EDotBox}), and  the addition formulas (\ref{d:addition}) we get
%The  map $\myPhi$, as well as $\tilde\myPhi$ is  either identical on those opposite faces in case a), or  antipodal in case b). We have that
\bge
\bga
c^3(\tilde\myPhi, \ct^2_{1-} \times E, S^2) - c^3( \myPhi, \ct^2_{1-} \times E, S^2) = 
\\
%\eps_{\al \be} \lb 
d \lb  \hkupt  , \hkup  , \ct^1_{3 --}\times E, S^2\rb
-  d \lb \hkupt  , \hkup , \ct^1_{3 + -}\times E, S^2\rb , 
%\rb
%\\
%- d \lb  \hkupt  , \hkup  , \ct^1_{\be, + \varsigma_\al }\times E, S^2\rb
%+ d \lb \hkupt  , \hkup , \ct^1_{\be, -\varsigma_\al}\times E, S^2\rb
\label{perC}
\ena
\ene
and similarly for $c^3(\tilde\myPhi, \ct^2_{2-} \times E, S^2)$.
%, which yields (\ref{additionT2})
Using periodic/antiperiodic conditions (\ref{perAper}), we get (\ref{additionT2}). 

\qed
%Since  $d_\be, d_\ga$ in (\ref{additionT2}) can be  any integers,  we have (\ref{HTsigma2A}) if $s_\be=s_\ga=0$, or (\ref{HTsigma2B})  if $s_\be=1$ or $s_\ga=1$.

\vspace{-3mm}
Define 
\bge
\dkk (\Phi_0,\Phi_1)= \mathop{\sum}_{\al=1}^2 c^3( \myPhi, \ct^2_{\al-} \times E, S^2) \ct^2_{\al-} .
\label{d2k}
\ene
\begin{thm}
Maps $\Phi_0$, $\Phi_1$ are homotopic if and only if $\dpst^1  (\Phi_0,\Phi_1)=0, \ \dpst^2 (\Phi_0,\Phi_1)=0$, and \\
1. Case (N) : $\dk (  \Phi_0, \Phi_1)=0$, $\dkk (  \Phi_0, \Phi_1)=0$ ,
\\
2. Case (T)  :  $\dk (  \Phi_0, \Phi_1) =0$, $\dkk (  \Phi_0, \Phi_1)=0 \ mod \  \mathop{\sum}_{\al,\be =1}^2  \eps_{\al \be} (1-(-1)^{s_\be})\ct^2_{\al-} 
%\ \  n\in \Bbb Z
$ .
\label{ThmPostHom}
\end{thm}
{\em Proof}.
It is clear that those conditions are necessary. Let us show that they are sufficient. Similarly to Theorem \ref{ThmConvexHom}, it follows from $\dpst^1  (\Phi_0,\Phi_1)=0, \ \dpst^2 (\Phi_0,\Phi_1)=0$ (\ref{d=0Post}) that there exist a surface homotopy $H_{\Phi_0,\Phi_1}^{2}$ on the truncated post surface. 

We will show that there exist a homotopy for maps restricted to $\kk$, the surface of the prism  bounding $\qqh$. It follows from $\dk (  \Phi_0, \Phi_1)=0$ that there exist a homotopy $\hk$ on edges of $\kk$. In case (N), it follows from  $\dkk (  \Phi_0, \Phi_1)=0$ and the periodic boundary conditions that there exist a homotopy $\hkk$ on the  vertical faces $\ct^2_{\al \pm}$, $\al=1,2$ of   $\kk$, extending edge homotopy $\hk$, $\hkk\vert(\kk_1\times E) = \hk\vert(\kk_1\times E)$. In case (T), we can select edge homotopy $\hk$ in such a way that we have $\dkk (  \Phi_0, \Phi_1)=0$, and thus there is a homotopy on vertical faces extending such $\hk$. 

On the top plate $\ct^2_{3  +}$, with $\myPhi$  defined in (\ref{Phi2EDotBox}), we always have  $ c^3(\myPhi, \ct^2_{3  +} \times E, S^2)=0$  due to the tangent boundary conditions, thus there is always a homotopy between restrictions of $\Phi_0$, $\Phi_1$ to the top plate. 

On the bottom plate, due to the periodic boundary conditions and the addition formula  (\ref{cISd}), (\ref{d:addition})  we have that 
$ c^2(\Phi_0, \ct^2_{3  -}, S^1)= c^2(\Phi_1, \ct^2_{3  -}, S^1)=0$, thus restrictions of  maps $\Phi_0$, $\Phi_1$ to the bottom plate have tangent extensions from the boundary of the  truncated post footprint   to its interior.  Thus as in the case of the top plate, there is always a homotopy between restrictions of $\Phi_0$, $\Phi_1$ to the   bottom plate. 

The above considerations yield a homotopy between restrictions of $\Phi_0$, $\Phi_1$ to the full boundary of $\qqh$ (consisting of the truncated surface of the post and a part of $\kk$ outside the truncated post).  The proof that the surface homotopy can be extended to the rest of $\qqh$ is the same as in Theorem \ref{ThmConvexHom}.

%\newpage

\end{document}